\documentclass[fleqn,10pt,table]{wlscirep}
\usepackage[utf8]{inputenc}
\usepackage[T1]{fontenc}
\usepackage{caption}
\usepackage{xcolor}
\usepackage{multirow}
\usepackage{amsmath,bm}
\usepackage{lineno}
\definecolor{Mycolor1}{HTML}{EBDEF0}
\definecolor{Mycolor2}{HTML}{d3dcf0}
\definecolor{c1}{HTML}{F9EBEA}
\definecolor{c2}{HTML}{F5EEF8}
\definecolor{c3}{HTML}{EAF2F8}
\definecolor{c4}{HTML}{E8F8F5}
\definecolor{c5}{HTML}{E9F7EF}
\definecolor{c6}{HTML}{FEF9E7}
\definecolor{c7}{HTML}{FDF2E9}
\definecolor{c8}{HTML}{F4F6F6}
\definecolor{c9}{HTML}{EAECEE}
\definecolor{c10}{HTML}{FEF5E7}
\definecolor{c11}{HTML}{FBEEE6}

\title{Evaluation of non-pharmaceutical interventions and optimal strategies for containing the COVID-19 pandemic}

\author[1,3*]{Xiao Zhou}
\author[2]{Xiaohu Zhang}
\author[3,4]{Paolo Santi}
\author[3]{Carlo Ratti}
\affil[1]{Gaoling School of Artificial Intelligence, Renmin University of China, Beijing 100872, China.}
\affil[2]{Department of Urban Planning, The University of Hong Kong, Hong Kong SAR, China.}
\affil[3]{Senseable City Laboratory, Massachusetts Institute of Technology, Cambridge, MA 02139, USA.}
\affil[4]{Istituto di Informatica e Telematica del CNR, Pisa, 56124, Italy.}

\affil[*]{corresponding.xiaozhou@ruc.edu.cn}

\begin{abstract}

Given multiple new COVID-19 variants are continuously emerging, non-pharmaceutical interventions are still primary control strategies to curb the further spread of coronavirus. However, implementing strict interventions over extended periods of time is inevitably hurting the economy. With an aim to solve this multi-objective decision-making problem, we investigate the underlying associations between policies, mobility patterns, and virus transmission. We further evaluate the relative performance of existing COVID-19 control measures and explore potential optimal strategies that can strike the right balance between public health and socio-economic recovery for individual states in the US. The results highlight the power of state of emergency declaration and wearing face masks and emphasize the necessity of pursuing tailor-made strategies for different states and phases of epidemiological transmission. Our framework enables policymakers to create more refined designs of COVID-19 strategies and can be extended to inform policy makers of any country about best practices in pandemic response.

\end{abstract}
\begin{document}

\flushbottom
\maketitle
%
%
\thispagestyle{empty}

\section*{Introduction}

Since the coronavirus disease 2019 (COVID-19) was initially detected in December 2019, a novel coronavirus designated as the severe acute respiratory syndrome coronavirus 2 (SARS-CoV-2) has rapidly spread across the world and led to a global pandemic \cite{world2020statement}. Despite increasingly more people getting vaccinated every single day, the world is still struggling to combat the emerging more contagious COVID-19 variants and has witnessed wave after wave of the epidemic here and there. As of 22 February 2022, a total number of 426,624,859 confirmed cases and 5,899,578 deaths due to the virus have been reported in 228 countries or territories globally according to the World Health Organization (WHO) \cite{WHOCovid19}. 

To contain the coronavirus outbreak, non-pharmaceutical interventions (NPIs) have been used widely as key weapons in some countries that were impacted heavily early on with satisfying effects \cite{flaxman2020estimating}. For instance, the world’s first stringent COVID lockdown sparked in Wuhan, the original epicenter of the pandemic. Specifically, China enacted travel bans to and from the city for 76 days and launched nationally coordinated efforts to mitigate the impact of the epidemic until it was eventually under control \cite{prem2020effect}. Similarly, in Italy, one of the European countries hardest hit by COVID-19, the national quarantine enforced on March 8, 2020, showed immediate effects on daily new infections within around two weeks \cite{CNN_2020}. Singapore is another country that adopted aggressive strategies and maintained a low casualty rate (0.15\%) compared to the global average (1.38\%) \cite{WHOCovid19}. Besides making striking achievements in real-world settings, the effectiveness of COVID-19 NPIs has also been extensively studied in the literature. Among them, a large body of research was targeted at the epidemiological implications of NPIs. Utilizing compartmental models, scholars simulated and predicted the effects of NPIs on multiple epidemic indicators, including infections \cite{davies2020effects,prem2020effect,cot2021mining,lai2020effect,flaxman2020estimating,ihme2020modeling,abueg2021modeling,singh2021impacts}, deaths \cite{lai2020effect,ragonnet2021genetic,flaxman2020estimating,ihme2020modeling,abueg2021modeling}, the reproduction number \cite{bo2021effectiveness,prem2020effect,flaxman2020estimating,ihme2020modeling}, and demand for hospital services \cite{davies2020effects,abueg2021modeling}. Touching on similar themes, another line of research focused more on quantifying the effects of NPIs on mobility \cite{hu2021big,askitas2021estimating,levin2021insights} and further explored the relationship between human movements and COVID-19 transmission \cite{cot2021mining,woskie2021early,oh2021mobility,ilin2021public}. Apart from characterizing the unfolding of the COVID-19 pandemic from an epidemiological perspective, there are also studies investigating the impact of NPIs from various angles covering economic contraction \cite{demi2020sooner,sheridan2020social,zaremba2020infected,demi2020sooner}, social issues \cite{zhang2020changes,abuhammad2021violence}, and mental health \cite{sutin2020change,luchetti2020trajectory,alzueta2021covid,sadikovic2020daily}. Some key findings summarized from the current state of knowledge include: (1) the imposition of appropriate NPIs is necessary to curb the spread of the virus \cite{flaxman2020estimating,oh2021mobility,ilin2021public}; (2) the effectiveness of individual intervention alone is usually limited and combined use of multiple NPIs is required \cite{ferguson2020report,lai2020effect} and; (3) fixed strategies of NPIs might be less effective than those with time-varying adjustment mechanisms \cite{nascimento2021reopening}. 

The encouraging precedents and research basis above demonstrate the role that NPIs can play in controlling the COVID-19 and offer valuable references for other affected regions to handle the spiraling outbreak. However, as time goes by, the situation has now become further complicated and highlighted some limitations of the current literature. First, the emphasis of many previous studies is on the epidemic simulation modeling of COVID-19 for more accurate forecasts without a thorough discussion of the relationships between human mobility and virus transmission that are deeply intertwined. In addition, given the core concern of handling the ongoing pandemic has shifted from containing the spread of the coronavirus only to restoring the social and economic order in the new normal from a long-term perspective, particular attention should be paid to multi-objective thinking in the design of policy strategies. Moreover, most of the above-mentioned successes were achieved by imposing harsh measures before vaccines became available, still comparatively little is known about how to devise policy frameworks that enable authorities to maintain a delicate balance between protecting public health and minimizing socio-economic disruption when the population are partly vaccinated against the COVID-19.

\begin{figure*}[h]
	\centering
	\includegraphics[width=16cm]{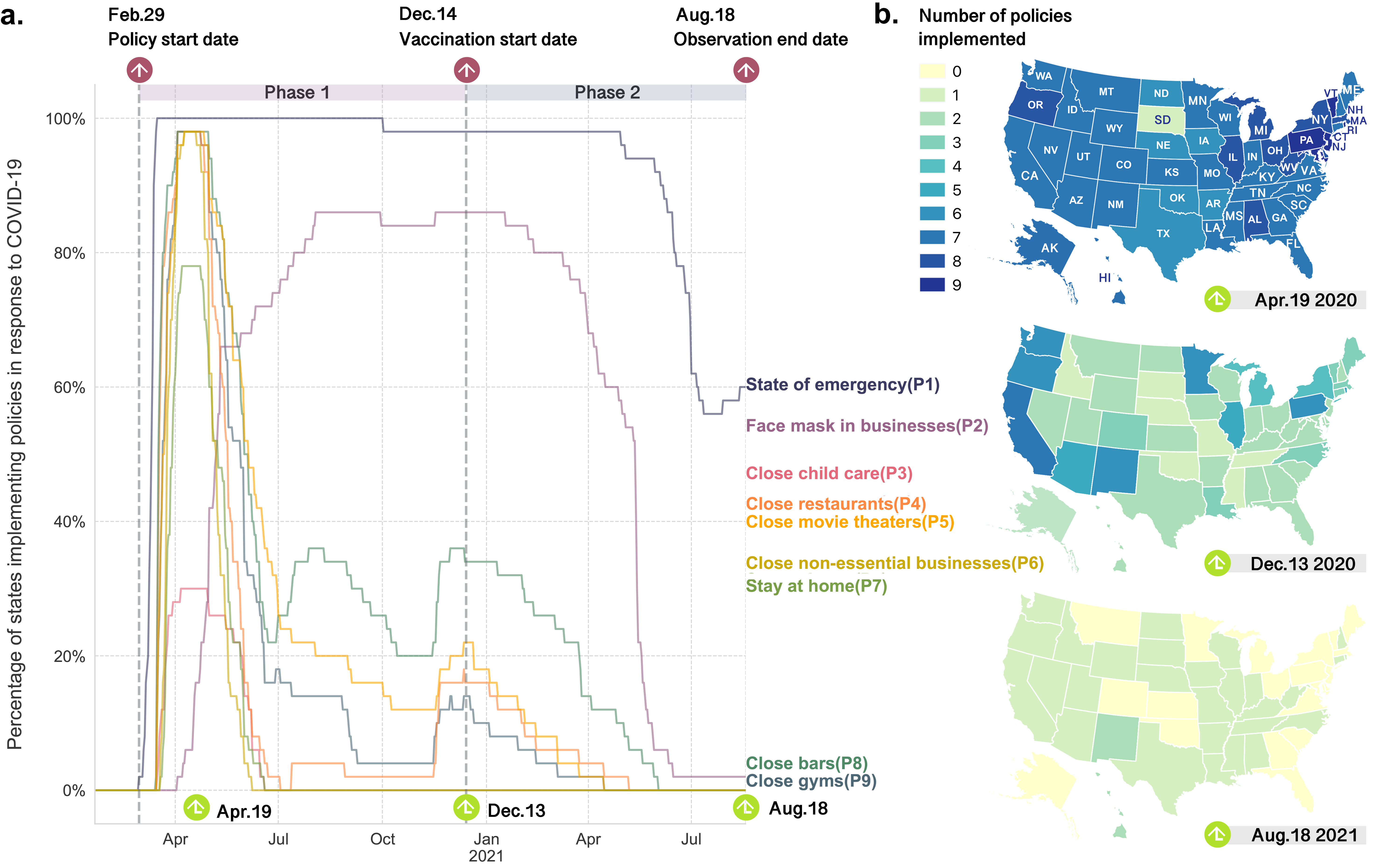}
	\caption{Implementation of anti-contagion policies in response to the COVID-19 over time and space in the United States. (\textbf{a}) Percentage of states with the COVID-19 policies being enacted over time. Two vertical dashed lines represent the start dates of COVID-19 policy deployment and vaccine distribution, respectively. (\textbf{b}) Maps of number of NPIs implemented by the states on three representative dates. April 19, 2020 represents the period with the most stringent lockdowns. December 13, 2020 is the day before COVID-19 vaccines were administered in the U.S.. August 18, 2021 is the last day of the observation period.}
	\label{fig1:policy_timeline}
\end{figure*}

To contribute in this direction, we choose the U.S. as the focus in the study due to its decentralized decision-making system, which leads to the implementation and enforcement of control measures highly variable in both time and space across the country with various circumstances for discussion. To get an overall picture of each state government’s NPIs in response to coronavirus, we employed the COVID-19 U.S. state policy Database \cite{raifman2020covid} and considered nine typical policies as listed in Table \ref{tab1:NPIs}. The adoption of these NPIs over time and space in the U.S. is presented in Fig. \ref{fig1:policy_timeline}. We further divide the observation period into two main phases according to when the lockdown restrictions and COVID-19 vaccine distribution started in the country. Some general patterns we find include: (1) most states rolled out the strictest lockdown measures in April 2020 to fight against the first wave and gradually rolled back the restrictions as the curve of infections started to flatten; (2) to control the subsequent waves in late June and mid-November 2020,
NPIs were reimplemented, however, the number of states participated is much smaller than previously and; (3) since the COVID-19 vaccines were administered, the NPIs were gradually lifted in most states. 

\begin{table}[ht]
\centering
\begin{tabular}{||c|c|c||}
\hline
\textbf{Label} & \textbf{Policy Name} & \textbf{Description} \\
\hline
\hline
\textbf{\textbf{P1}}& State of emergency & A state issued any type of emergency declaration \\
\hline
\textbf{\textbf{P2}} & Face mask in businesses & Mandate face mask use by employees in public-facing businesses \\
\hline
\textbf{\textbf{P3}}& Close child care & A state closed day cares statewide \\
\hline
\textbf{\textbf{P4}} & Close restaurants & A state closed restaurants (except for take out) \\
\hline
\textbf{\textbf{P5}} & Close movie theaters & A state closed movie theaters statewide \\
\hline
\textbf{\textbf{P6}}& Close non-essential businesses & A state closed non-essential businesses statewide \\
\hline
\textbf{\textbf{P7}}& Stay at home & A state issued statewide stay at home/shelter in place order \\
\hline
\textbf{\textbf{P8}} & Close bars & A state closed bars statewide \\
\hline
\textbf{\textbf{P9}} & Close gyms & A state closed indoor gyms/fitness centers \\
\hline
\end{tabular}
\caption{\label{tab1:NPIs} Summary of typical NPIs considered and corresponding descriptions.}
\end{table}

Using the COVID-19 policy dataset along with epidemiological and mobility data aggregated at the state level, we propose a methodological framework to quantify the underlying relationships between mobility and key measures of population-level transmission in the context of the COVID-19 pandemic both before and after the vaccines became available in the U.S.. We further evaluate the effectiveness of current combinations of NPIs applied in individual states and identify new optimal strategies that can well balance the public health and socio-economic impacts in the fight against coronavirus. Finally, we investigate the similarities and differences across states and phases to provide new insights into the spatio-temporal dynamics for epidemic control. The findings enable policymakers to better understand the rapidly evolving nature of the pandemic and signpost optimal and more flexible solutions that can adapt to changing needs and circumstances.

\section*{Results}

\subsection*{Temporal associations between COVID-19 transmission, population mobility, and state policies}

\begin{figure*}[h]
	\centering
	\includegraphics[width=17cm]{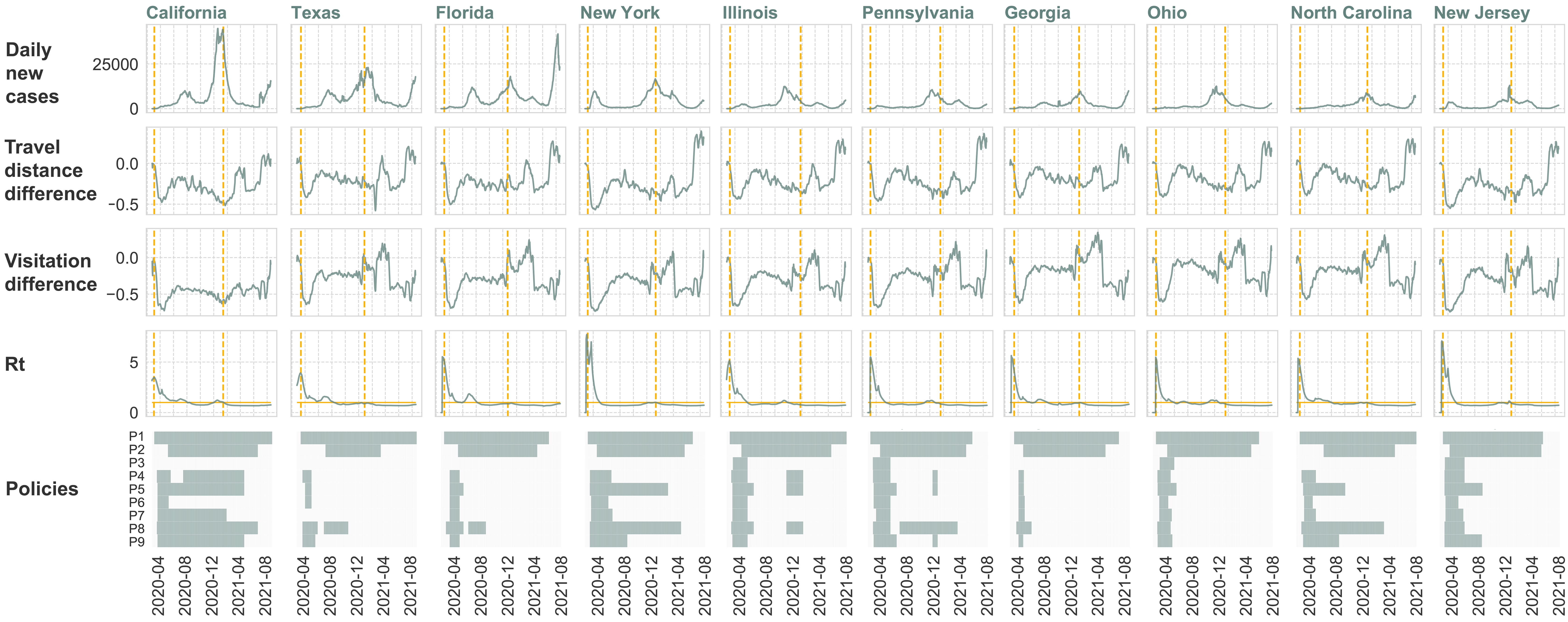}
	\caption{Temporal changes in daily new COVID-19 cases, travel distance difference, visitation difference, instantaneous reproduction number $R_t$ and policy implementation in the ten states with highest number of confirmed cases from February 24, 2020 to August 18, 2021. 7-day moving average is utilized to smooth volatile case reporting data and human mobility metrics. The start dates of policy implementation and vaccine distribution are indicated by dashed vertical lines. A horizontal line is drawn at $R_t$=1. If $R_t$ is greater than 1, the epidemic is expanding at time $t$, whereas $R_t<1$ signals that the epidemic is shrinking. }
	\label{fig2:Timeline_m}
\end{figure*}

Apart from the database of NPIs, we also employ daily COVID-19 case data from the New York Times \cite{NYT2021} and aggregated mobility data from Unacast \cite{Unacast_mobility} to draw a comprehensive comparison of the temporal variations in virus transmission, mobility patterns, and policies during the COVID-19 outbreak. Here, two mobility metrics utilized are relative percentage changes in visits to non-essential venues and average travel distance compared to the corresponding day of week prior to the COVID-19 outbreak for a given date. We map the timelines of the state government policy actions against the daily new reported cases, human mobility trends, and instantaneous reproduction number $R_t$ estimated by the susceptible-exposed-infectious-recovered-susceptible (SEIRS) epidemic model \cite{bjornstad2020seirs}(see the Methods) in Fig. \ref{fig2:Timeline_m}. Here, $R_t$ is a key indicator to monitor the real-time transmissibility of the virus and to estimate the impact of local NPIs on the epidemic. 

As Fig. \ref{fig2:Timeline_m} suggests, the state authorities generally started to take measures at the exact time when the highest $R_t$ appeared and employed the NPIs again when the coronavirus spikes were reaching alarming levels during the subsequent waves. One can also notice that the mobility level seems to be an indicator of how stringent the local NPIs are. In the beginning, when the states adopted the strictest interventions, the population mobility experienced a sharp decline from peak to trough in all the states. Then with the adjustments of the NPIs and the application of vaccines, the relative reduction in mobility also changed accordingly. In terms of the relationships between new infections and human mobility, the relevant curves in Fig. \ref{fig2:Timeline_m} hint at the existence of time-lagged associations, which will be further investigated through vector autoregression model \cite{sims1980macroeconomics} and Granger causality analysis \cite{Granger:1969} next. Moreover, other intertwined relationships between NPIs, mobility patterns, and the spread of the virus that we have a glimpse into here will also be explored using multiple linear regression later. Due to space limitations, we only display the detailed analysis results of California and key findings of the other nine states severely affected by the pandemic as examples in the main text. Additional findings of other states in the U.S. are presented in Supplementary Information.

\subsection*{Temporal lagged relationships between mobility and viral transmissibility}

\begin{figure*}[h]
	\centering
	\includegraphics[width=15cm]{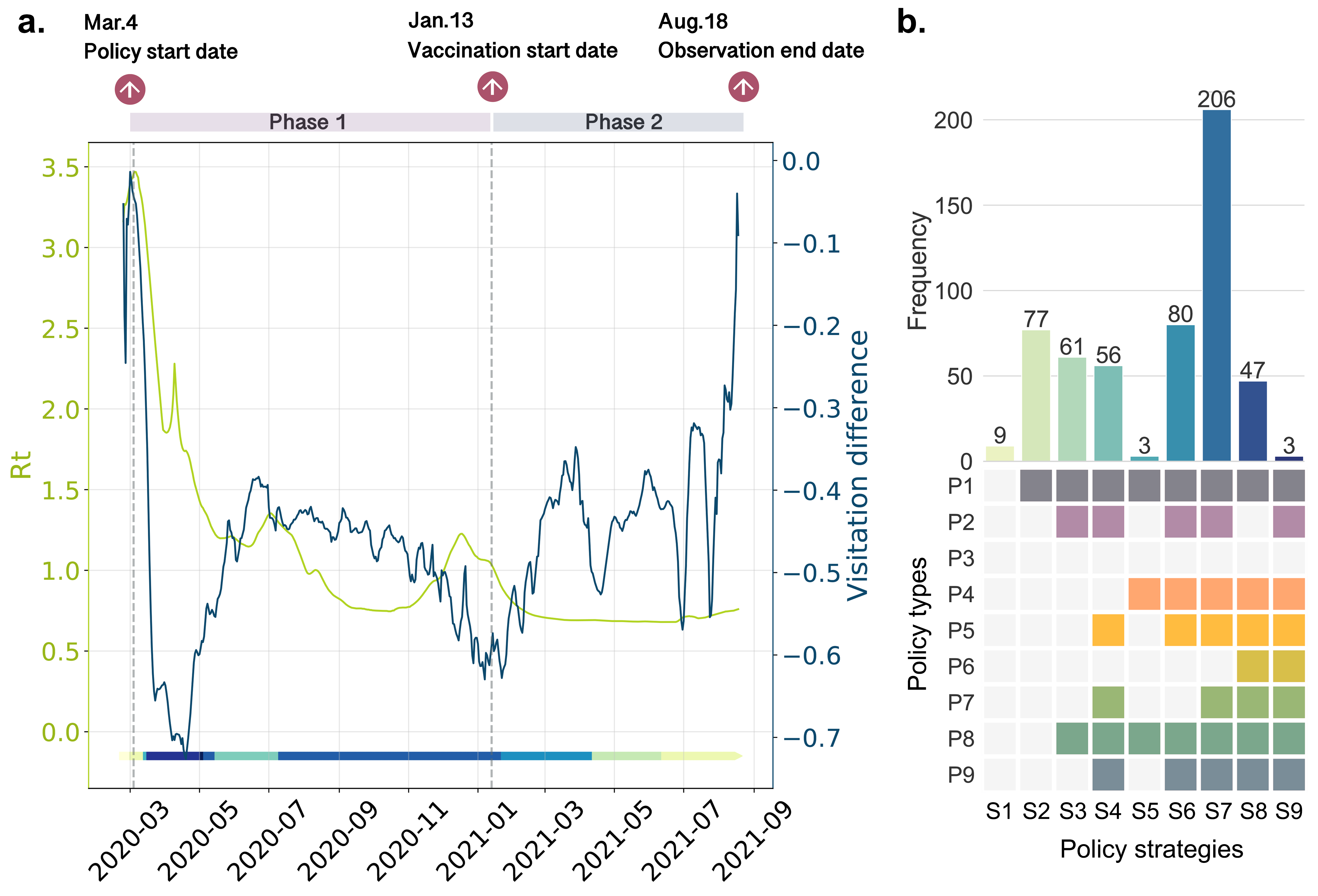}
	\caption{Temporal changes in $R_t$, \textit{VD}, and COVID-19 policy implementation in California during two phases. The start dates of policy implementation and vaccine distribution are indicated by dashed vertical lines. Existing policy sets and their frequencies are displayed in (\textbf{b}). The temporal distribution of the policy sets is presented on the bottom of (\textbf{a}). The curves of $R_t$ and \textit{VD} are perceived as more tortuous in phase 1. After the vaccines were released, \textit{VD} roughly showed a rising trend and $R_t$, in contrast, declined gradually and kept steady below 1 with a slight increase displayed toward the end of observation.}
	\label{fig3:CA_RtV}
\end{figure*}

Prior to examining the dynamic linkages between human mobility and virus transmission, we display basic descriptive statistics in Table \ref{tab2:Descriptive_statistics} and time series of variables selected in Fig. \ref{fig3:CA_RtV}. Here we employ a sub-period analysis to explore how variables and the relationships between them changed over time before and after the administration of the COVID-19 vaccines. Table \ref{tab2:Descriptive_statistics} and Fig. \ref{fig3:CA_RtV} present that $R_t$ in California experienced a significant decrease from up to 3.469 in phase 1 to below 1 during phase 2. In terms of the mobility changes, a clear pattern emerged is that California suffered relatively greater declines in travel distance and visits during phase 1 compared to its counterpart. In addition, as Fig. \ref{fig3:CA_RtV} depicts, the strategies adopted during phase 1 are generally more diverse and stricter than those in phase 2. A visual inspection in Fig. \ref{fig3:CA_RtV} suggests that $R_t$ and \textit{VD} show opposite trends in general with temporal lagged associations and present different tendencies before and after the advent of the vaccines.

\begin{table}[ht]
\centering
\begin{tabular}{||c|c|c|c|c|c|c|c|c|c|c||}
\hline
 &\multicolumn{5}{c|}{\textbf{Phase 1 (Mar.4 2020 - Jan.12 2021)}} & \multicolumn{5}{c||}{\textbf{Phase 2 (Jan.13 2021 - Aug.18 2021)}}\\
\cline{2-11}
\textbf{Variables} & \textbf{\textit{NC}} & \textbf{\textit{ND}} & \cellcolor{Mycolor1} $\bm{R_t}$ & \textbf{\textit{TD}} & \cellcolor{Mycolor1}\textbf{\textit{VD}} & \textbf{\textit{NC}} & \textbf{\textit{ND}} & \cellcolor{Mycolor2}$\bm{R_t}$ & \textbf{\textit{TD}} & \cellcolor{Mycolor2}\textbf{\textit{VD}}\\
\hline
\hline
\textbf{Mean} & 8934 & 99 & 1.255 & -0.309 & -0.489 & 6503 & 157 & 0.721 & -0.217 & -0.437 \\
\hline
\textbf{Min} & 10 & 0 & 0.746 & -0.485 & -0.726 & 883 & 16 & 0.680 & -0.517 & -0.628 \\
\hline
\textbf{Max} & 44768 & 531 & 3.469 & -0.024 & -0.044 & 39587 & 561 & 1.021 & 0.117 & -0.040 \\
\hline
\textbf{Std.Dev.} & 11780.600 & 90.258 & 0.588 & 0.102 & 0.110 & 7253.021 & 169.436 & 0.062 & 0.182 & 0.091 \\
\hline
\end{tabular}
\caption{\label{tab2:Descriptive_statistics}Summary statistics for COVID-19 and mobility variables in California during different phases. Phase 1 represents the period when NPIs have been adopted but no vaccines had become available yet, while phase 2 represents the time when the NPIs and vaccines were deployed together. \textit{NC} represents daily new reported cases. \textit{ND} means daily new deaths. Mobility variables of \textit{TD} and \textit{VD} represent the changes in average travel distance and visits compared to those for the same day of week during non-COVID-19 time period, respectively.}
\end{table}

Based on the preliminary observations, we explore multiple temporal dynamic relationships between transmission and mobility variables in a bivariate setting by using vector autoregressive models \cite{sims1980macroeconomics} and the Granger causality tests \cite{Granger:1969}. For clarity, we will take the tests between $R_t$ and \textit{VD} as an example to illustrate in detail how the techniques are employed below. 

Vector autoregression (VAR) \cite{sims1980macroeconomics} is a widely used statistical method for multivariate time series analysis. Granger causality analysis \cite{Granger:1969} was initially developed in econometrics as a technique for investigating the directed interactions between time-series data \cite{stokes2017study}. This statistical concept of causality is based on the prediction that a time series \textbf{\textit{x}} (\textit{VD}) Granger causes another time series \textbf{\textit{y}} ($R_t$) if the autoregressive forecast of \textbf{\textit{y}} can be better explained when the past information from \textbf{\textit{x}} is considered \cite{papagiannopoulou2017non}. After determining the maximum order of integration ($d$) and optimal time lag length ($m$) for $R_t$ and \textit{VD} (see Supplementary Table 1-2), we establish bivariate augmented VAR models for the two phases in California, based on the idea of the Toda-Yamamoto Granger causality test \cite{toda1995statistical} (see the Methods) as follows:

\begin{equation}\label{eq2}
y_t = \gamma +\sum_{i=1}^{m+d}\alpha_{i}y_{t-i}+\sum_{i=1}^{m+d}\phi_{i}x_{t-i}+\varepsilon_{t}
\end{equation}

\noindent where $y_t$ denotes the value of the time series \textbf{\textit{y}} at time $t$, $i$ is the length of the lag-time moving window, $\alpha_i$ and $\phi_i$ are the parameters to estimate, $\epsilon_t$ refers to the white noise residual. The variables of \textbf{\textit{x}} and \textbf{\textit{y}} can be interchanged to test for the Granger causality in the other direction.

The results in Table \ref{tab3:GC_CA} suggest that the Granger causality relationship from $R_t$ to $VD$ is statistically significant at the 1\% level during phase 1 in California, but not significant for phase 2. The phenomena can be interpreted as people reacting to the news about confirmed coronavirus cases by changing their mobility patterns significantly in phase 1. This behavioral response, however, is not that evident after the vaccines were available, hinting to the fact that people felt more protected by vaccines and less keen on constraining their movements. To look in the other direction, it is found that the null hypothesis of no Granger causality from $VD$ to $R_t$ can be rejected at the 10\% significance level during phase 2, which implies that the visitation change of individuals in California Granger causes the transmission of the COVID-19 during the period of phase 2. 

Apart from the variable pair of $R_t$ and $VD$, the Toda and Yamamoto causality test is also performed between other possible pairs of the epidemiological and mobility variables following this framework. It can be observed from Table \ref{tab3:GC_CA} that bi-directional Granger causality emerged between variable pairs of ($ND$, $TD$) in both phases and ($ND$, $VD$) in phase 2. Moreover, all the test results for $ND$ during phase 2 reject the null hypothesis at the significance level of 1\%. For the corresponding tests between $NC$ and the two mobility variables, no Granger causality relationship is found at the 1\% significance level. This observation of stronger Granger causality relationships existing between $ND$ and mobility patterns compared to those of $NC$ seems to be consistent with the research findings of some previous papers \cite{ku2020epidemiological, flaxman2020report,hadjidemetriou2020impact}, in which death count was considered as a more reliable metric over daily infections, since the actual number of infected cases is expected to be significantly larger than what has been reported. It is also noticeable that the optimal lag lengths selected according to information criteria for $ND$ and two mobility metrics are higher than those of $NC$ of around one to two weeks, which may correspond to the length of treatment.

\begin{table}[ht]
\centering
\begin{tabular}{||c|c|c|c|c|c|c|c|c||}
\hline
 &\multicolumn{4}{c|}{\textbf{Phase 1 (Mar.4 2020 - Jan.12 2021)}} & \multicolumn{4}{c||}{\textbf{Phase 2 (Jan.13 2021 - Aug.18 2021)}}\\
\cline{2-9}
\textbf{Direction} & \textbf{Lag(m)} & \textbf{Lag(m+d)} & \textbf{Chi-square} & \textbf{Prob.} & \textbf{Lag(m)} & \textbf{Lag(m+d)} & \textbf{Chi-square} & \textbf{Prob.}\\
\hline
\hline
$\bm{R_t}$ $\rightarrow$ \textbf{\textit{VD}} & 2 & 2 & $16.086^{***}$ & \cellcolor{Mycolor1}{0.000} & 9 & 10 & 11.680 & \cellcolor{Mycolor2}{0.307} \\
\hline
\textbf{\textit{VD}} $\rightarrow$ $\bm{R_t}$ & 2 & 2 & 2.531 & \cellcolor{Mycolor1}{0.282} & 9 & 10 & $16.209^{*}$ & \cellcolor{Mycolor2}{0.094} \\
\hline$\bm{R_t}$ $\rightarrow$ \textbf{\textit{TD}} & 2 & 3 & 4.544 & 0.208 & 8 & 9 & 5.687 & 0.771 \\
\hline
\textbf{\textit{TD}} $\rightarrow$ $\bm{R_t}$ & 2 & 3 & 1.449 & 0.694 & 8 & 9 & 5.520 & 0.787 \\
\hline
\hline
\textbf{\textit{NC}} $\rightarrow$ \textbf{\textit{VD}} & 9 & 10 & 4.565 & 0.918 & 8 & 9 & $19.714^{**}$ & 0.020 \\
\hline
\textbf{\textit{VD}} $\rightarrow$ \textbf{\textit{NC}} & 9 & 10 & 9.258 & 0.508 & 8 & 9 & $16.648^{*}$ & 0.055 \\
\hline
\textbf{\textit{NC}} $\rightarrow$ \textbf{\textit{TD}} & 8 & 9 & $18.044^{**}$ & 0.035 & 8 & 9 & 7.689 & 0.566 \\
\hline
\textbf{\textit{TD}} $\rightarrow$ \textbf{\textit{NC}} & 8 & 9 & 10.722 & 0.295 & 8 & 9 & 4.660 & 0.863 \\
\hline
\hline
\textbf{\textit{ND}} $\rightarrow$ \textbf{\textit{VD}} & 16 & 17 & 13.924 & 0.672 & 23 & 24 & $51.655^{***}$ & 0.001 \\
\hline
\textbf{\textit{VD}} $\rightarrow$ \textbf{\textit{ND}} & 16 & 17 & 15.914 & 0.530 & 23 & 24 & $46.339^{***}$ & 0.004 \\
\hline
\textbf{\textit{ND}} $\rightarrow$ \textbf{\textit{TD}} & 23 & 24 & $35.936^{*}$ & 0.056 & 20 & 21 & $57.509^{***}$ & 0.000 \\
\hline
\textbf{\textit{TD}} $\rightarrow$ \textbf{\textit{ND}} & 23 & 24 & $46.526^{***}$ & 0.004 & 20 & 21 & $58.966^{***}$ & 0.000 \\
\hline
\end{tabular}
\caption{\label{tab3:GC_CA}Toda-Yamamoto Granger causality test results for transmission and mobility variable pairs in California. ***, **, and * indicate the rejection of the null hypothesis at the 1\%, 5\% and 10\% significance levels, respectively.}
\end{table}

To check if the VAR models are well specified, we also conducted a series of statistical tests (See Supplementary). The underlying temporal associations between the spread of coronavirus and mobility patterns discovered in this subsection will be considered in the following Pareto analyses for the assessment and optimal design of the NPIs.

\subsection*{Pareto optimality for COVID-19 policy assessment and design.}


In multi-objective optimization problems, the Pareto-efficient state is achieved if there is no other solution that can bring improvement to one of the objectives without showing degradation in another objective \cite{ngatchou2005pareto}. According to this, our multi-objective optimization problem can be defined as a vector function \textbf{\textit{f}} that maps a vector of policy decision variables \textbf{\textit{p}} to a tuple of two objectives \textbf{\textit{h}} as follows:

\begin{equation}
\begin{aligned}
  \text{minimize: } &\textbf{\textit{f}}(\textbf{\textit{p}}) = min \left \{ w_1f_1(\textbf{\textit{p}}), w_2f_2(\textbf{\textit{p}})\right \}\\
  \text{subject to: } &\textbf{\textit{p}} = (p_1,p_2,...,p_k) \in P \\
  &\textbf{\textit{h}} = (h_1,h_2) \in H
\end{aligned}
\end{equation}

\noindent where $k$ is the number of policy types, $P$ is the policy space, and $H$ is the objective space.

In our case, the optimization goal is to strike a delicate balance between the control of the COVID-19 virus and the recovery of socio-economic vitality, which are indicated specifically by $R_t$ and \textit{VD}, respectively. These two variables are chosen because we consider the estimated $R_t$ a more comprehensive metric to measure the transmissibility of the epidemic and visitation frequency a more appropriate indicator of socio-economic vitality. Based on this assumption, one of the objectives in our research task is to minimize the reproduction number of the virus $f_r(\textbf{\textit{p}})$. In the meanwhile, the value of visitation metric $f_v(\textbf{\textit{p}})$ is expected to be maximized. Accordingly, we set the values of $w_1$ and $w_2$ to 1 and -1. Consider two policy decision vectors $\textbf{\textit{a}}, \textbf{\textit{b}} \in P$. The policy decision vector $\textbf{\textit{a}}$ is said to dominate $\textbf{\textit{b}}$ if their objective vectors $\textbf{\textit{f}}(\textbf{\textit{a}})$ and $\textbf{\textit{f}}(\textbf{\textit{b}})$ satisfy:

\begin{equation}
\begin{aligned}
  && f_r(\textbf{\textit{a}}) \leqslant f_r(\textbf{\textit{b}}) \wedge f_v(\textbf{\textit{a}}) > f_v(\textbf{\textit{b}}),\\
or && f_v(\textbf{\textit{a}}) \geqslant f_v(\textbf{\textit{b}}) \wedge f_r(\textbf{\textit{a}}) < f_r(\textbf{\textit{b}})
\end{aligned}
\end{equation}

\noindent Under such rules, a decision vector is said to be Pareto optimal if and only if there does not exist another solution that dominates it. The set of all the policy decision vectors that are not dominated by any other generates the Pareto optimal set, while the corresponding objective vectors are said to be on the Pareto frontier \cite{ngatchou2005pareto}. Identifying the Pareto frontier is particularly useful because it can provide policymakers with a group of optimal solutions to make a well-informed decision that balances the trade-offs between the public health concerns and socio-economic losses rather than a single-point solution. 

\subsubsection*{Evaluation of existing COVID-19 policy strategies.}

We first use the notion of Pareto optimality to evaluate the performance of existing policy combinations. Specifically, for each given date $t$, we collect the corresponding value of visitation metric $v_t$ and estimate the reproduction number ${R}_t^{'}$ using the SEIRS epidemic model with temporal lagged effects considered (see Methods). These two features collectively form the two-dimensional space, in which Pareto-optimal set would be explored. For each Pareto-optimal point obtained, we can figure out the corresponding control measures implemented on a particular date for further investigation. The following offers the results of optimal solutions discovered using this approach in the actual scene.

\begin{figure*}[h]
	\centering
	\includegraphics[width=16cm]{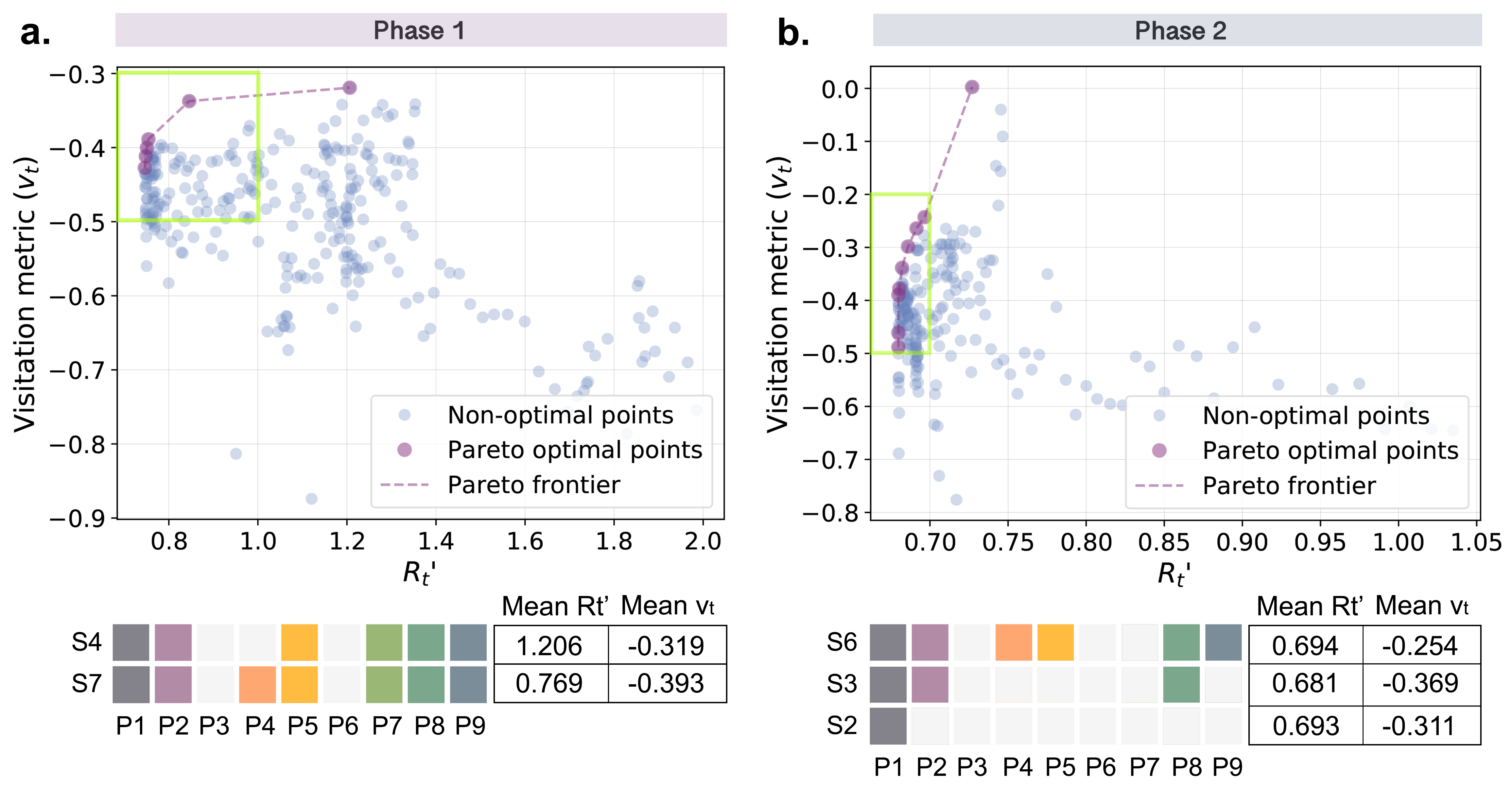}
	\caption{Pareto optimal trade-offs between human mobility ($v_t$) and virus transmission (${R}_t^{'}$) in California during two phases. Purple-colored spots represent optimal solutions that are connected by dashed lines to visually estimate the Pareto frontier. Candidate points with a value of ${R}_t^{'}$ larger than 2 are filtered. Pareto optimal points with a ${R}_t^{'}$ between 0.7 and 1 are enclosed by green boxes for phase 1 and phase 2, respectively. Corresponding optimal policy strategies after duplicate elimination are displayed in lower sub-figures with average ${R}_t^{'}$ and $v_t$ listed in the tables beside them.}
	\label{fig4:Pareto1_CA}
\end{figure*}

As can be observed from Fig. \ref{fig4:Pareto1_CA}, there are in total two and three unique Pareto optimal solutions for California in phase 1 and phase 2, respectively. Among them, the state of emergency (P1) is found to be the only individual policy present in all the optimal solutions. To discuss by phases, the combined power of stay-at-home order (P7), the closure of movie theaters (P5), gyms (P9), bars (P8), and wearing face masks in businesses (P2) is particularly strong when pending the availability of vaccines since they showed together in both of the optimal policy sets during phase 1. The closure of restaurants (P4) is the only policy type that distinguishes strategy S7 from S4. It is interesting to observe that the closure of restaurants is connected to a considerable reduction in $R'_{t}$, down to 0.769 from 1.206, confirming the relatively high risks of transmitting the virus during convivial activities such as dining in a group. However, closing restaurants also considerably reduces the mobility index, down to -0.393 from -0.319, confirming the trade-off between the need of containing the virus and socio-economic vitality. However, when it came to phase 2, a larger number of enacted control measures did not guarantee a smaller ${R}_t^{'}$. Instead, the lowest average ${R}_t^{'}$ is generated by a moderate policy strategy S3, where mask mandate is in place and only bars are closed. In addition, the slope of the Pareto frontier for phase 2 is greater than that for phase 1 in California, implying that the increase in ${R}_t^{'}$ is accompanied by a relatively larger recovery of $v_t$ when the vaccines became available. 

It should also be noticed that the solutions selected are considered equally good according to the Pareto optimality concept. To decide which solution to choose depends on the policy makers’ perspectives about the priority of the two objectives in the optimization task. For instance, if the decision-makers in California intend to relax policies to some extent so that they have as little as possible impact on the normal mobility of the residents in phase 1, the corresponding ${R}_t^{'}$ would be as high as 1.206 on average as solution S4 presents.

\subsubsection*{Design of optimal control strategies for the COVID-19 pandemic.}

Following the assessment of the existing policies, we explore possible new policy combinations that might be more effective than the current ones next by employing the non-dominated sorting genetic algorithm II (NSGA-II) \cite{deb2002fast} (see Methods). 

Specifically, we adopt multiple regression (see Methods) first to estimate the coefficient for each policy type in the prediction of $\hat{R}_t$ and $\hat{v}_t$ during different phases with temporal lags effects considered. These parameters calculated are then fed into the NSGA-II algorithm to generate the Pareto optimal solutions. Then the solutions with an estimated $\hat{R}_t$ no larger than the maximum mean ${R}_t^{'}$ of the existing optimal policy strategies are further selected. Different from the implementation of existing policies that are dummy-coded, the parameter estimated through the generic algorithm for each type of policy is a continuous variable, which can be interpreted as the strength of the policy enforcement, where a value closer to 1 represents the strength is relatively stronger, whereas closer to 0 indicates that the implementation of the policy is weaker. 

\begin{figure*}[h]
	\centering
	\includegraphics[width=17cm]{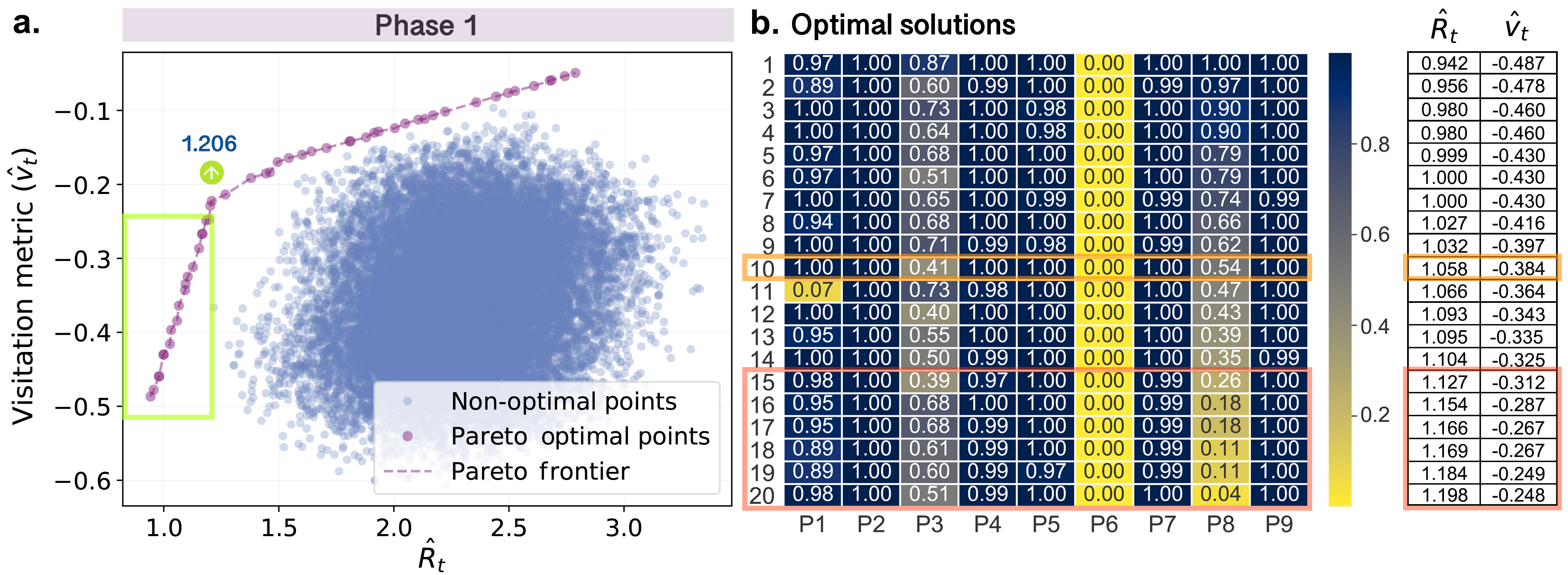}
	\caption{Potential Pareto optimal policy strategies generated for California in phase 1. (\textbf{a}) visually presents the Pareto solutions. Spots and dashed line in purple represent the analytically derived Pareto optimal set and frontier. Optimal points with a $\hat{R}_t$ less than 1.206 are selected. Corresponding parameters estimated for the policy types are displayed in (\textbf{b}) with predicted $\hat{R}_t$ and $\hat{v}_t$ listed in the right table. Orange rectangle highlights the solution similar to the existing strategy of S7; and, coral ones highlight more optimized solutions. Predicted values of the two objectives are summarized in the table.}
	\label{fig5:Pareto2_CA_PnV}
\end{figure*}

The optimization results for California during phase 1 is presented in Fig. \ref{fig5:Pareto2_CA_PnV}. Comparing these optimization results with those of the existing policy strategies, the last six potential solutions are found more optimized than S4 since they have both smaller $\hat{R_t}$ and larger $\hat{v_t}$. This observation suggests that exploring new possible strategies for better trade-offs between virus control and the maintenance of mobility is necessary. Among theses six optimal solutions, the NPIs of state emergency declaration (P1), wearing face masks in businesses (P2), closing restaurants (P4), movie theaters (P5), and gyms (P9), and issuing stay-at-home order (P7) are expected to be implemented together with strong strengths. In contrast, the closure of non-essential businesses (P6) seems to be not very necessary, which is in accordance with the phenomenon observed in the evaluation of existing optimal strategies previously. The main distinctions between the potential optimal solutions discovered here from the existing ones for phase 1 lie in the adoption of P3 and P8. Essentially, the new proposed optimal solutions put more emphasis on the importance of closing child care centers and the flexible adjustment of the shutdown of bars. 
 
\begin{figure*}[h]
	\centering
	\includegraphics[width=17cm]{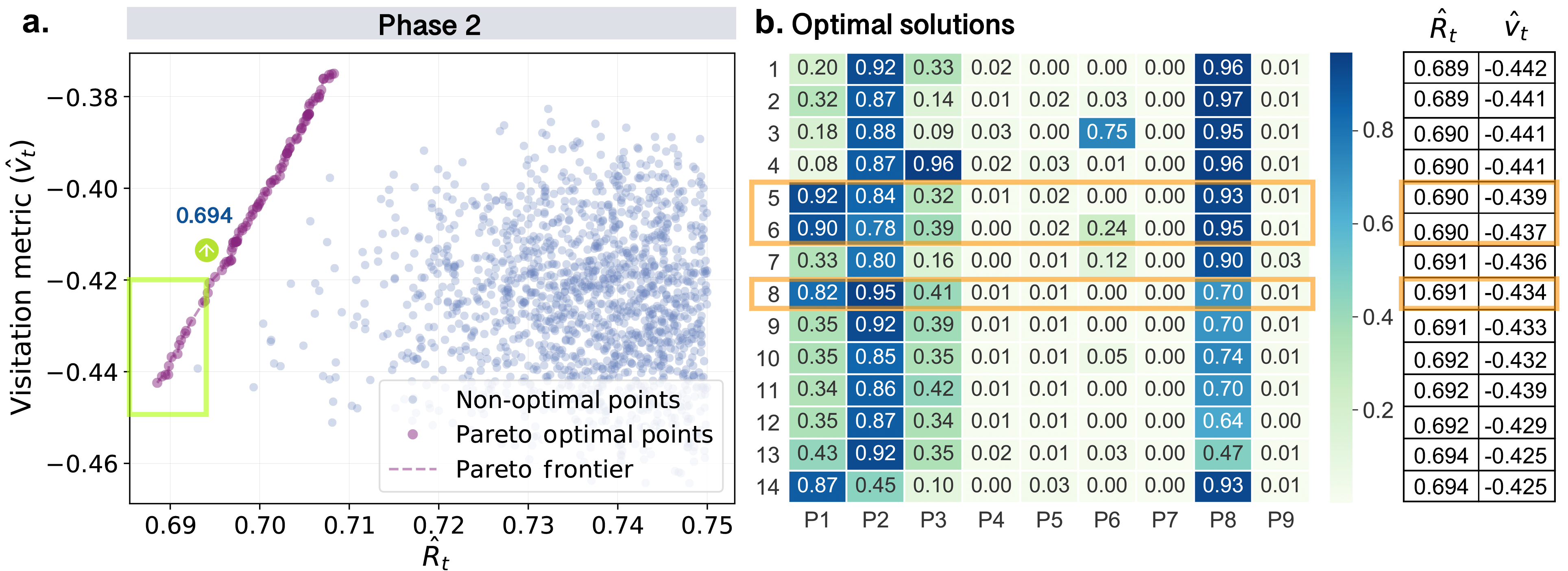}
	\caption{Potential Pareto optimal policy strategies generated for California in phase 2. (\textbf{a}) visually presents the Pareto solutions. Optimal points with a $\hat{R}_t$ no more than 0.694 are selected. Corresponding parameters estimated for the policy types are displayed in (\textbf{b}) with predicted $\hat{R}_t$ and $\hat{v}_t$ listed in the right table. Orange rectangles highlight the solution similar to the existing strategy of S3.}
	\label{fig6:Pareto2_CA_PV}
\end{figure*}

While for the design of optimal strategies for California in phase 2 (Fig. \ref{fig6:Pareto2_CA_PV}), it presents a significantly different pattern from that for phase 1. Specifically, the overall strength of the policies for phase 2 is relatively weaker, and the solutions generated are more heterogeneous. In addition, policymakers in California should also shift the focus of the policy types to wearing face masks in businesses (P2) and closing bars (P8) in phase 2. In the meantime, the closures of restaurants (P4), movie theaters (P5), gyms (P9), and the implementation of the stay-at-home order (P7) during this period are considered less significant than those during phase 1. 

\begin{figure*}[h]
	\centering
	\includegraphics[width=17cm]{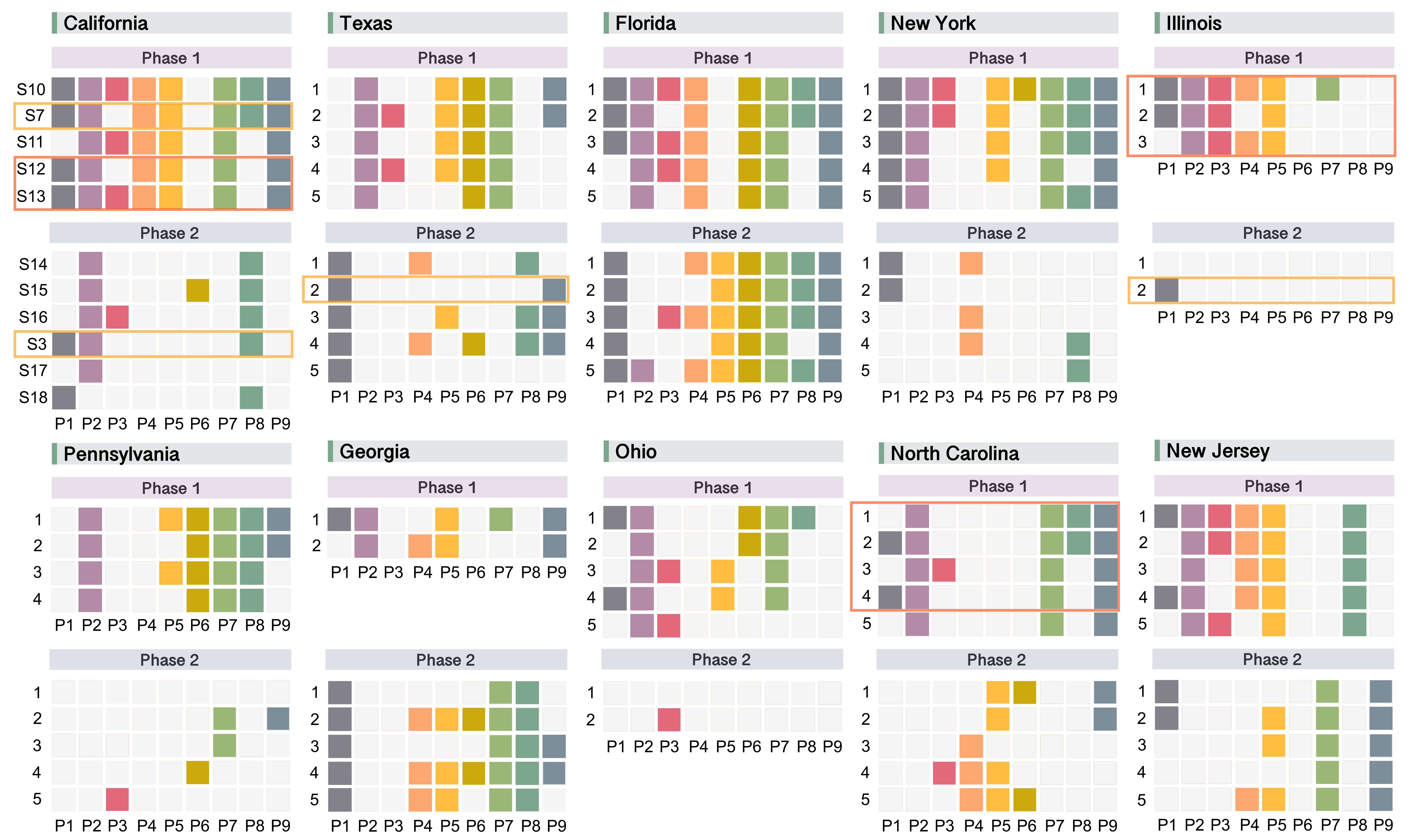}
	\caption{Optimal response strategies generated by NSGA-II for ten states with the highest number of confirmed COVID-19 cases during different phases. Strategies for each state are listed in ascending order of the average predicted $\hat{R_t}$. If more than five optimal solutions are distilled for a certain state and phase, only the top five strategies are retained. Orange rectangles mark solutions already included in the existing strategies for the state; and, coral rectangles highlight more optimized ones.}
	\label{fig7:Pareto2_filter_all}
\end{figure*}

To provide a clearer picture of how to sketch out the implementation plans, we then replace policy parameters that fall within the ranges (0, 0.5) and (0.5, 1) as 0 and 1, respectively. This simplified version of the optimal policy scheme is shown in Fig. \ref{fig7:Pareto2_filter_all}, from which we can see that there are five and six types of optimal policy strategies extracted for phase 1 and phase 2 in California, respectively. Among them, the current solutions of S7 and S3 are included, while the other nine potential optimal strategies that may achieve better or equally good performance compared to the existing ones offer decision-makers a list of possible alternatives in the fight against coronavirus. 

Following the same framework, we conduct analyses for other states and display their simplified optimal strategies in Fig.\ref{fig7:Pareto2_filter_all}. From the view of comparisons across space and time, following major findings emerge: (1) the differences between optimal policy designs for the ten states studied are evident, indicating the necessity of adopting differentiated and tailor-made COVID-19 response strategies in each individual state; (2) existing policy strategies (enclosed by orange rectangles) seldom appear in the list of new proposed ones, suggesting that there are a considerable number of alternatives available for COVID-19 prevention beyond current policy programs; (3) new solutions generated that are more preferable than the current ones (enclosed by coral rectangles with smaller $\hat{R_t}$ and higher $\hat{v_t}$) show in the first phase of some states (California, Illinois, and North Carolina), offering potential solutions that could be considered to replace the existing ones to policymakers; (4) for each state, two different phases present widely divergent strategies regarding policy types and amount of enacting policies. Specifically, the designs for phase 2 include fewer policy types and amounts compared to phase 1 in general, the phenomenon of which may be explained by the availability of vaccines in phase 2 when the control of COVID-19 no longer relies solely on the NPIs.

\section*{Discussion}


In this paper, we propose a flexible decision support system that allows state policymakers to evaluate and create refined COVID-19 policy strategies based on the art of balancing public health and socio-economic vitality. Here tailor-made adjustment schemes and novel optimal policies in response to the COVID-19 pandemic are generated for state authorities to choose depending on their priorities and current vaccination status. This can be achieved through the joint utilization of epidemiological data, mobility data, NPIs, and in-depth exploration of the dynamic relationships between them by fusing multiple techniques. 

We started with a discussion of the intertwined associations among the spread of the virus, policy implementation, and human mobility during different phases and discovered some prevailing patterns across the states (Fig. \ref{fig2:Timeline_m} and Supplementary Fig. 2-5). For instance, the state governments generally began to implement NPIs when the $R_t$ was at its maximum. In addition, human mobility can be used as a proxy for NPIs that 
it dropped steeply to the lowest level when the most stringent NPIs were issued in the early stages of the pandemic. Then with the subsequential relaxation and tightening of NPIs later on, human mobility levels also changed accordingly. The emergence of COVID-19 vaccines introduced new variables into these interrelationships. Some states (e.g., CA, IL, PA, MI, MN, WA, IA, and OR) began to loosen their NPIs, which coincided with virus transmission decline and human mobility increase.

Another key observation that warrants further investigation is the existence of underlying temporal associations between human mobility and virus transmission, which might be bidirectional and dynamic.
For instance, people would adjust their travel behaviors if they saw a dramatic rise in coronavirus cases; and the changes in mobility trends would, in turn, affect the confirmed cases in a few days. 
The results of VAR models prove the existence of temporal lagged effects as well as Granger causality relationships between some transmission and mobility variable pairs during different phases (Table \ref{tab3:GC_CA}). With regard to the variable pair of ($R_t$, $VD$) that we pay particular attention to, significant Granger causality relationships are largely discovered in the states affected the most by coronavirus in the U.S. (Supplementary Table 4). Here for each state, at least one direction in one phase offers results that are significant at the 5\% level. These findings help deepen our understanding of the transmission dynamics of COVID-19 and human behaviors by uncovering the intertwined temporal associations between them, and thus enable more holistic policy-making with temporal effects taken into account.

We then evaluate the performance of existing policy strategies across the states using Pareto analysis, so as to offer policymakers a lens for looking back to identify effective policy combinations and adjust less effective ones in the future. For the pre-vaccine stage, to measure the contribution of policy types to health-economic balance individually, state of emergency (P1) and wearing face masks (P2) show as the most important NPIs that they appear in all the optimal strategies among the top ten states (Supplementary Table 5). The closure of recreational services (P8, P5, P9, P4) comes next with moderate impact, while essential businesses (P6) and child care centers (P3) do not present in any of the optimal strategies. From the perspective of the policy mix, the co-activation of P1 and P2 accounts for the largest proportion (30\%) and is present in the optimal strategies for six states (TX, FL, IL, GA, OH, NJ). The second most frequent policy combination is (P1, P2, P8), which accounts for 25\% of the optimal strategies. 
This finding implies that paying closer attention to the control of gathering sites where social scenes are more complex (e.g., bars) is particularly necessary before the advent of vaccines. Then when the COVID-19 vaccines were available, the percentage of each policy type shown in the optimal strategies has decreased in general (Supplementary Table 6). However, from a ranking view, what does not change compared to phase 1 is that the most effective individual policy types are still P1, P2, and P8. 

Finally, we employ NSGA-II algorithm to generate new optimal policy strategies for each state and phase. Here, the results reveal substantial differences between different phases and states. For instance, wearing face masks (P2) is particularly important for all the states in the pre-vaccine period (Supplementary Fig. 7-15). However, this is not the case in the later phase. The policy type of closing restaurants (P4) is essential for FL and NJ in phase 1, but not that important for the states of TX, NY, PA, and NC. In addition, new optimal solutions created differ significantly from the existing ones. In a nutshell, these findings illustrate the necessity of designing tailor-made new policy strategies through genetic algorithms for individual states and phases. The results obtained here can provide policymakers in each state with a general guiding policy formulation reference. It also enables them to compare the expected effects of potential optimal solutions with the performance of existing strategies and make more deliberate decisions regarding what kind of adjustments shall be made to the current schemes and how. Moreover, the implementation strategies proposed by our framework are spatio-temporal, context-aware, and more fine-grained than the existing policy schemes. This is reflected in the state-tailored considerations of temporal lag effects, Granger causality relationships, policy-based transmission and mobility predictions in a phase-wise manner. Furthermore, our framework also offers sophisticated implementation strategies that produce the strengths of policy enforcement as outputs for decision-makers.

Our study has several limitations. First, we mainly focus on NPIs implementation in the regions with the most confirmed cases, since they play a more crucial role in combating the COVID-19 pandemic. However, an in-depth analysis of areas with fewer cases may raise other intriguing questions, such as why they performed relatively better in the fight against coronavirus. This sort of exploration may lead to a deeper understanding of how to improve the combined power of NPIs or offer novel insights into reasons for the effectiveness of control measures varies considerably across regions. This future investigation may require the additional use of demographic data, which can also help reveal the underlying logic of the results we have obtained. In addition, we perform  COVID-19 policy-making optimization by considering two primary objectives in the study. However, many other aspects also need particular attention in the decision-making process in real-world settings, like the psychological distress and economic recession following the COVID-19 outbreak.


In conclusion, we provide detailed insights into the spatio-temporal dynamics of the COVID-19 epidemic during different stages and highlight the essential role of some core intervention portfolios in the controlling of the pandemic. The methodology proposed here can offer policymakers reasoned estimates of the potential effectiveness of the NPIs to be attained and becomes even more critical when health systems are facing extreme loads. Furthermore, for other populations lacking local COVID-19 data and in the future similar circumstances for other emerging infectious diseases in which an outcome of interest is not yet available, the framework presented can also play a part. 

\section*{Methods}

\subsection*{Epidemiological model.}

We apply a modified SEIRS \cite{bjornstad2020seirs} compartmental model to detect the local spread of COVID-19 in the U.S. at different stages and estimate the effects of containment measures on the epidemic evolution in each state. 

In epidemiology, classical compartmental models provide a simplified representation of the mechanisms of virus propagation by dividing the population into different subgroups according to the infection status from the transmission process. Depending on the epidemiological characteristics and transmission patterns of infectious diseases, appropriate epidemic models should be chosen accordingly. Most notably, a canonical model used in this area is the SIR model \cite{kermack1927contribution}, in which individuals are assumed to be in one of three distinct epidemiological phases: susceptible (\textbf{\textit{S}}), infected (\textbf{\textit{I}}), and recovered (\textbf{\textit{R}}). \textbf{\textit{S}} represents a group of individuals who have not yet been infected by the virus; \textbf{\textit{I}} models infected people with symptoms of the disease who can spread the virus to individuals in the \textbf{\textit{S}} compartment; and, \textbf{\textit{R}} stands for those who have recovered from the disease and become immune to the virus. Considering the existence of incubation period, the SEIR model adds an exposed state (\textbf{\textit{E}}) between susceptibility and infection to the basic SIR model to represent individuals who have contracted the virus, showing no visible clinical signs of the disease, and may infect others. In recent decades, the SEIR model has been widely used for simulating the spread of infectious diseases, including severe acute respiratory syndrome (SARS) in 2002 \cite{lloyd2003curtailing}, H1N1 influenza in 2009 \cite{bajardi2011human}, and the 2014 Ebola virus outbreak in West Africa \cite{althaus2014estimating}. SEIR model has also been recently adopted in the analysis of the COVID-19 pandemic \cite{prem2020effect,peirlinck2020outbreak,linka2020outbreak}. The SEIR model assumes individuals carry lifelong immunity to the disease upon recovery, however, there have been clinical findings showing that patients who have recovered from COVID-19 can get reinfected with the virus \cite{radulescu2020management,ng2020covid,to2020covid}. Even though the immune response to this novel virus is not yet fully understood, the possibility of reinfection cannot be ruled out. To reflect this, we adopt a modified SEIRS model tailored for COVID-19 instead, allowing recovered individuals to return to the susceptible state after some time when their immunity vanishes. Moreover, an additional compartment representing the number of individuals died due to the virus is added in our model and is denoted by \textbf{\textit{D}}. It is the only compartment that would be removed from the total population \textbf{\textit{N}} and have no further interactions with the rest of the epidemic system. The coupled dynamics of the compartments are governed by the following set of ordinary differential equations:

\begin{equation}
\frac{dS(t)}{dt} = -\frac{\beta S(t)I(t)}{N(t)}+\xi R(t)
\end{equation}

\begin{equation}
\frac{dE(t)}{dt} = \frac{\beta S(t)I(t)}{N(t)}-\sigma E(t)
\end{equation}

\begin{equation}
\frac{dI(t)}{dt} = \sigma E(t)-(\gamma+ \mu)I(t)
\end{equation}

\begin{equation}
\frac{dR(t)}{dt} = \gamma I(t) -  \xi R(t)
\end{equation}

\begin{equation}
\frac{dD(t)}{dt} = \mu I(t)
\end{equation}

\noindent where $\beta$ is the rate of infection transmission, which is normalized by $N(t)$, representing the total population at time $t$: $N(t) = S(t)+E(t)+I(t)+R(t)$. $\sigma$ denotes the incubation rate of latent individuals becoming symptomatic, and is calculated as the inverse of incubation period length of COVID-19. Similarly, $\gamma$ represents the recovery rate, the inverse of which is the average time an infected person needs to be recovered. In our model, patients are assumed to develop short-term immunity of $1/\xi$ days after recovering from the viral infection and become susceptible to the virus again. However, their unlucky counterparts who died from the virus at the fatality rate $\mu$ would be removed from the transmission process spontaneously. 

In compartmental epidemic models, a key parameter used to characterize the transmissibility of the virus is called the basic reproduction number ($R_0$). It corresponds to the average number of secondary cases generated by per infectious individual in a fully susceptible population and can be computed from the following equation:

\begin{equation}
R_{0}=\frac{\beta }{\gamma + \mu }
\end{equation}

\noindent While the $R_0$ is typically considered as a biological and clinical characteristic of a virus measuring how contagious the infectious disease is, it is also influenced by NPIs. The logic behind this is that a drastic reduction in mobility associated with the implementation of COVID-19 containment measures would decrease the contact rate between infected and susceptible individuals, leading to a decline in the estimated $\beta$ and $R_0$. Under this assumption, a sliding window-based extension of the SEIRS model is proposed to capture the real-time epidemic dynamics of the Coronavirus during different phases of intervention development. Specifically, using a sliding window of size $2n+1$, we calculate a time-varying transmission rate of the infection at time $t$ as:

\begin{equation}
{\beta(t)}' =\sum_{i = t-n}^{2n+1}\beta(i)/(2n+1)
\end{equation}

Based on the dynamic contact rate $\beta(t)'$, we then infer the instantaneous reproduction number $R_t$ to track the epidemic progression over time and estimate the impact of local interventions. Here, if $R_t$ is greater than 1, the epidemic is expanding at time $t$, whereas $R_t<1$ indicates that the epidemic is shrinking. The expression for the $R_t$ is:

\begin{equation}
R_t =\frac{\beta(t)'}{\gamma + \mu }
\end{equation}

\noindent It allows rapid detection of the ongoing evolution during the COVID-19 pandemic, taking the virus's epidemiological characteristics and human mobility patterns that may be significantly affected by local control measures into account. Specifically, we simulate the local spread of the COVID-19 pandemic in different states using the compartmental model introduced above, fitted with publicly available data from the New York Times \cite{NYT2021}. Disease-specific parameters in the model were derived from the recent literature, including a mean incubation period of 5.2 days (95\% confidence interval [CI] 4.1-7.0) \cite{li2020early}, an average recovery time of 8 days \cite{maier2020effective} and the mean time to death from onset of 17.8 days (95\% CI 16.9-19.2) \cite{verity2020estimates}. In addition, patients are assumed to develop temporal immunity of $\xi = 180$ days, after recovering from the initial infection of the virus \cite{seow2020longitudinal}. 

\subsection*{Granger causality analysis.}

To introduce the idea of Granger causality more concretely, let \textbf{\textit{x}} $= [x_1, x_2, …, x_n]$ and \textbf{\textit{y}} $= [y_1, y_2, …, y_n]$ denote the two stationary time series of length $n$. To examine whether time series \textbf{\textit{x}} Granger causes \textbf{\textit{y}}, consider the following model:

\begin{equation}\label{eq1}
y_t = \gamma +\sum_{i=1}^{m}\alpha_{i}y_{t-i}+\sum_{i=1}^{m}\phi_{i}x_{t-i}+\varepsilon_{t}
\end{equation}

\noindent where $y_t$ denotes the value of the time series \textbf{\textit{y}} at time $t$, $i$ is the length of the lag-time moving window, $\alpha_i$ and $\phi_i$ are the parameters to estimate, $\epsilon_t$ refers to the white noise residual. In this setting, the Granger causality can be investigated based on an $F$ test with the null hypothesis of $H_0: \phi_1 = \phi_2 = ... = \phi_i = 0 $. Essentially, if $H_0$ is rejected, one can conclude that the Granger causality from time series \textbf{\textit{x}} to \textbf{\textit{y}} exists, since that the lagged values of \textbf{\textit{x}} provide additional explanatory and predictive power to the regression model. Based on the conventional Granger causality test introduced above, Toda and Yamamoto \cite{toda1995statistical} proposed a modified version that overcomes several limitations of traditional approaches in hypothesis testing when there are unit roots in the VAR system. Specifically, Toda and Yamamoto causality test does not require pre-testing for the cointegrating properties of the system, and thus avoids the potential bias associated with unit roots and cointegration tests \cite{clarke2006comparison, zapata1997monte}. Given these advantages of the Toda and Yamamoto causality test, we adopt this method in the present study.

\subsection*{Multiple linear regression analysis.}

Multiple linear regression analysis is used to assess the association between policy types and estimate regression coefficients. For each state, we build two multiple linear regression models, in which independent variables are policy types ${[p_1, p_2, …, p_k]}$, while the continuous dependent variables are $v_{t}$ and ${R_{t}}^{'}$, respectively. Here,  ${R_{t}}^{'}$ denotes the temporal lagged reproduction number for date $t$ with individualized temporal relationships between the $v_{t}$ and $R_{t}$ in each state considered. The assumption here is that the changes in human mobility are visible on the same date when the NPIs are issued, while ${R_{t}}^{'}$ is determined by the temporal lagged associations between mobility and the spread of the coronavirus investigated using VAR model with Granger causality test. For instance, since there is a significant Granger causality relationship discovered from $R_{t}$ to $v_{t}$ with a lag of $m=2$ in California during phase 1, the corresponding ${R_{t}}^{'}$ for $v_{t}$ is equal to $R_{t-2}$. This operation applies to both of the Pareto analysis parts, including the evaluation of existing policy strategies and the design of new policy solutions. To accomplish the second task, the processed data is used to fit two multiple linear regression models for each phase in a state as follows:

\begin{equation}\label{eq1}
{R_t}^{'} = {\lambda_0}+{\lambda_1}{p_1}+{\lambda_2}{p_2}+\cdots+{\lambda_k}{p_k}
\end{equation}

\begin{equation}\label{eq1}
{v_t} = {\eta_0}+{\eta_1}{p_1}+{\eta_2}{p_2}+\cdots+{\eta_k}{p_k}
\end{equation}

\noindent The estimated parameters of ${[\lambda_0, \lambda_1, \lambda_2, …, \lambda_k]}$ and ${[\eta_0, \eta_1, \eta_2, …, \eta_k]}$ are used for the prediction of $\hat{R_t}$ and $\hat{v_t}$ in the generation process of new optimal policy strategies.

\subsection*{Pareto optimality.}

The concept of Pareto efficiency \cite{pareto1964cours} is originally introduced to describe an economic state in which the reallocation of resources cannot make at least one person better off without making any other individual worse off \cite{Fudenberg1991}. To generate new policy strategies for each individual state, we employ the NSGA-II algorithm \cite{deb2002fast} by using the Platypus package \cite{hadka2019platypus} for multi-objective evolutionary computing in Python 3.7. 

NSGA-II differs from traditional genetic algorithms in two aspects: (1) the selection of appropriate individual solutions among possible ones to generate the next generation is based on their dominance levels according to the Pareto optimality; (2) The crowding distance is employed as a measure to make choices between individual solutions that have the same dominance level. Additionally, NSGA-II adopts an elitist strategy for the selection that unless better solutions are found, the best ones obtained so far are retained. The algorithm produces a set of optimal solutions that collectively make up the Pareto front when the optimization process is terminated \cite{yoo2010using}.

\section*{Data Availability.}

All data used in this manuscript are publicly available. COVID-19 epidemiological data are available from the New York Times at [https://github.com/nytimes/covid-19-data]\cite{NYT2021}. Daily aggregated mobility data at the state level are available for each state from Unacast [https://www.unacast.com/covid19/social-distancing-scoreboard]\cite{Unacast_mobility}. State-wide policies related to COVID-19 are collated by the researchers at the Boston University School of Public Health [https://github.com/USCOVIDpolicy/COVID-19-US-State-Policy-Database]\cite{raifman2020covid}. 

\section*{Code availability.}

The code and processed data to reproduce tables and figures in the main text and Supplementary Information are publicly available on GitHub at: https://github.com/XiaoZHOUCAM/COVID-NPIs-Optimal.

\bibliography{sample}



\section*{Author contributions}

X.Z.,X.H.Z, and P.S. conceived the experiments. X.H.Z collected the pandemic and mobility data. X.Z. conducted the experiments. P.S. and C.R. coordinated the project. X.Z. wrote the paper. All authors participated in the discussion of results, provided critical feedback and reviewed the manuscript. 

\section*{Competing Interests}

The authors declare no competing interests.

\end{document}